\documentclass[prd,showkeys,floatfix,twocolumn,amsmath,amssymb,floatfix]{revtex4}
\usepackage{graphicx,color,dcolumn,booktabs,bm}
\usepackage{longtable,lscape}
\usepackage{amssymb}
\usepackage{amsmath}
\usepackage{indentfirst}
\usepackage{epsfig}
\usepackage{feynmf}
\usepackage{subfigure}
\usepackage{epstopdf}
\usepackage{slashed}
\usepackage{cases}
\usepackage{multirow}
\usepackage[colorlinks,citecolor=blue,anchorcolor=red,menucolor=red,linkcolor=red,filecolor=red,runcolor=red,urlcolor=blue,frenchlinks=red]{hyperref}

\allowdisplaybreaks[3]

\begin{document}

\title{QCD Sum Rule Analysis of Topped Mesons in the Heavy-Quark Limit}

\author{Shu-Wei Zhang$^1$}
\author{Xuan Luo$^1$}
\author{Hui-Min Yang$^{1,2}$}
\author{Hua-Xing Chen$^1$}
\email{hxchen@seu.edu.cn}

\affiliation{$^1$School of Physics, Southeast University, Nanjing 210094, China\\
$^2$School of Physics and Center of High Energy Physics, Peking University, Beijing 100871, China}

\begin{abstract}
Motivated by the recent CMS observation of a near-threshold enhancement in top quark pair production, we investigate a novel class of hadronic systems containing a single top quark: the topped mesons (\( t\bar{q} \), with \( \bar q = \bar u, \bar d, \bar s \)). In contrast to the extensively studied toponium (\( t\bar{t} \)) system—analyzed primarily within perturbative QCD—topped mesons offer a complementary nonperturbative probe of QCD dynamics in the heavy quark limit. These states are expected to exhibit longer lifetimes and narrower decay widths than toponium, as only a single top quark undergoes weak decay. We employ QCD sum rules within the framework of heavy quark effective theory to study the structure and mass spectrum of ground-state topped mesons. Our analysis predicts masses near 173.1~GeV, approximately 0.5–0.6~GeV above the top quark pole mass. Compared with singly topped baryons (\( tqq \), with \( q = u, d, s \)), topped mesons have a simpler quark composition and more favorable decay channels ({{a} 
topped meson is anticipated to decay weakly into a $\Upsilon$ meson and a charmed meson}), enhancing their potential for both theoretical analysis and experimental discovery.
\end{abstract}


\maketitle

\pagenumbering{arabic}

\section{Introduction}

The top quark holds a unique status within the Standard Model due to its remarkably large mass of approximately 172.57~GeV~\citep{pdg}, setting it apart from all other quark flavors. Unlike lighter quarks, it decays before hadronization can occur, owing to its extremely short lifetime of about \( 5 \times 10^{-25} \) s. This exceptional characteristic presents both theoretical challenges and novel opportunities for the study of hadronic physics. Theoretical efforts have traditionally focused on the \textit{toponium} system—a hypothetical \( t\bar{t} \) bound state. Although such a state cannot form a stable resonance, due to the rapid decay of the top quark, the dynamics of near-threshold \( t\bar{t} \) production are still governed by nonrelativistic QCD. Extensive studies based on perturbative QCD and potential models have predicted distinctive enhancements near the production threshold~\citep{Fadin:1987wz,Barger:1987xg,Kuhn:1987ty,Fadin:1990wx,Strassler:1990nw,Sumino:1997ve,Hoang:2000yr,Penin:2005eu,Hagiwara:2008df,Kiyo:2008bv,Sumino:2010bv,Beneke:2015kwa,Fuks:2021xje,Wang:2024hzd,Jiang:2024fyw,Akbar:2024brg,Fuks:2024yjj}. Notably, the CMS Collaboration recently reported an excess in the \( t\bar{t} \) invariant mass spectrum near the threshold~\citep{CMS:2025kzt}, consistent with the expected behavior of a pseudoscalar toponium state. Similar features observed in earlier ATLAS and CMS data further support this interpretation~\citep{ATLAS:2023fsd,CMS:2024pts}. The observed enhancement appears more consistent with QCD-based toponium predictions than with alternative explanations, potentially marking the first experimental evidence of toponium formation—an important milestone in top quark physics~\citep{Aguilar-Saavedra:2024mnm,Llanes-Estrada:2024phk,Fu:2025yft,Fu:2025zxb,Nason:2025hix,Ellis:2025nkm,Jafari:2025rmm,Bai:2025buy,Shao:2025dzw,Xiong:2025iwg,Thompson:2025cgp}.

Beyond toponium, a broader class of hadronic systems containing a single top quark—such as \textit{topped mesons} and \textit{singly topped baryons}—presents additional theoretical interest~\cite{liuxiang}. Although it is not so easy for these configurations to form bound states, due to the top quark’s fleeting lifetime, they provide a valuable framework for probing heavy quark dynamics. A particularly noteworthy feature lies in their decay properties: since only one top quark decays, their expected lifetimes are approximately twice that of toponium, resulting in narrower decay widths~\cite{Chen:2021erj,Maltoni:2024csn}. {Therefore, if toponium can exist as a physical state, it is reasonable to expect that hadronic systems containing a single top quark may also manifest under suitable conditions. These characteristics may enhance their experimental accessibility and open new avenues for exploring top-related QCD phenomena.}

The study of topped mesons and singly topped baryons can be systematically carried out within the framework of heavy quark effective theory (HQET)~\cite{Neubert:1993mb,Manohar:2000dt}. In this approach, the top quark is treated as a static color source, enabling a systematic classification of the light degrees of freedom. This simplification facilitates the construction of interpolating currents suitable for both QCD sum rule analyses and lattice QCD simulations~\cite{Dai:1993kt,Liu:2007fg}. Singly topped baryons are thoroughly examined in Ref.~\cite{Zhang:2025xxd}, while the present work focuses on a systematic investigation of topped mesons. In this study, we construct and analyze interpolating currents for ground-state topped mesons within the HQET framework. Using QCD sum rules, we explore their properties and find that their predicted masses lie near 173.1~GeV—approximately 0.5–0.6~GeV above the top quark pole mass.

Since topped mesons have not been previously studied, no established theoretical framework currently exists for their description. The application of HQET and QCD sum rules in this context remains speculative and is not fully justified:
\begin{itemize}

\item The lifetime of topped mesons is significantly shorter than their formation time, causing them to behave as extremely wide resonances in top quark collisions. While HQET may be applicable in this regime due to its assumption of infinite heavy quark mass, this assumption still requires further justification.

\item QCD sum rules rely on the separation between the resonant region—where correlators can be approximated by pole saturation of a few low-lying resonances—and the higher-mass continuum. For extremely wide resonances such as topped mesons, the distinction between the continuum and the ``resonant region'' may become invalid and also requires further justification.

\end{itemize}

Therefore, we interpret our results as exploratory estimates rather than definitive predictions. The primary goal of this work is to probe the theoretical limits of nonperturbative methods in this novel regime, providing insight into the potential behavior of topped mesons. We also emphasize that the predictions made in this study are testable. Should the experimental results confirm these predictions, it would represent a breakthrough in understanding these systems. Conversely, if the predictions are not supported by experimental data, it would point out areas where these methods need further refinement and improvement, guiding future theoretical research into the nature of these novel \mbox{hadronic states.}

It is worth emphasizing that topped mesons offer several distinct advantages over their baryonic counterparts. First, their internal structure is conceptually simpler and more transparent: only two categories are expected, namely pseudoscalar mesons ($T$ and $T_s$) and vector mesons ($T^*$ and $T_s^*$). The states \( T^{(*)} \) and \( T_s^{(*)} \) correspond to the quark configurations \( t \bar{q} \) (with \( \bar q = \bar u \text{ or } \bar d \)) and \( t \bar{s} \), respectively. Second, a topped meson involves only a single light antiquark, making it generally easier to produce than a singly topped baryon, which requires two light quarks. Third, a topped meson is expected to decay weakly into a $\Upsilon$ meson and a charmed meson—final states that are typically easier to identify and reconstruct experimentally than those from a singly topped baryon, which decays into a $\Upsilon$ meson and a charmed baryon.

This paper is organized as follows. In Section~\ref{sec:leading} we construct interpolating currents for ground-state topped mesons and perform QCD sum rule analyses at leading order. In Section~\ref{sec:nexttoleading} we incorporate \(\mathcal{O}(1/m_Q)\) corrections to examine next-to-leading-order effects. Finally, Section~\ref{sec:summary} summarizes our findings and discusses their physical implications.

\section{Results at Leading Order}
\label{sec:leading}

In this section we perform leading-order QCD sum rule analyses to investigate the properties of ground-state topped mesons. Before presenting the formalism, we briefly outline the notations and conventions adopted throughout this work.

A topped meson is defined as a bound state composed of a heavy top quark and a light antiquark, which may be of up, down, or strange flavor. The lowest-lying topped mesons carry spin-parity quantum numbers \( J^P = 0^- \) and \( 1^- \), and are organized into a doublet under heavy quark effective theory (HQET). This structure arises in the heavy quark limit, where the top quark decouples from the system’s dynamics, leaving the light degrees of freedom—characterized by the light-quark spin \( s_q = 1/2 \)—to govern the quantum behavior. Coupling the light component with the heavy quark spin \( s_Q = 1/2 \) yields two degenerate states with total angular momentum \( J = s_Q \otimes s_q = 0 \oplus 1 \), forming an HQET doublet.

The relevant interpolating currents are constructed {as}
\begin{eqnarray}
J(x) &=& \bar q^{a}(x) \gamma_5 h_v^a(x) \, , 
\label{def:J0}
\\
J_\mu(x) &=& \bar q^{a}(x) \gamma^t_\mu h_v^a(x) \, ,
\label{def:J1}
\end{eqnarray}
where \( a \) denotes the color index. The field \( \bar q(x) \) refers to a light antiquark, such as \( \bar u(x) \), \( \bar d(x) \), or \( \bar s(x) \), while \( h_v(x) \) represents the effective top quark field in HQET. The transverse gamma matrix is defined by \( \gamma^t_\mu = \gamma_\mu - v\!\!\!\slash\, v_\mu \), where \( v_\mu \) is the four-velocity of the \mbox{heavy quark.}

Since \( J(x) \) and \( J_\mu(x) \) arise from the same HQET doublet, they yield equivalent QCD sum rules. Thus, it is sufficient to work with either current. In this analysis, we focus on the vector current \( J_\mu(x) \) with quark content \( t \bar{s} \) to investigate the properties of the vector topped meson \( T_s^\star \), which carries quantum numbers \( J^P = 1^- \). The corresponding decay constant is defined through the matrix element:
\begin{equation}
\langle 0 | J_\mu(x) | T_s^\star \rangle = f \, \epsilon_\mu \, ,
\end{equation}
where \( f \) denotes the decay constant and \( \epsilon_\mu \) is the polarization vector. This leads to the following two-point correlation function:
\begin{eqnarray}
\Pi_{\mu\nu}(\omega) &=& i \int d^4 x\, e^{i k \cdot x} \langle 0 |
T[J_\mu(x) J_\nu^\dagger(0)] | 0 \rangle
\label{eq:pi}
\\ \nonumber
&=& \left(g_{\mu\nu} - \frac{q_\mu q_\nu}{q^2} \right) \times \Pi(\omega) + \cdots \, ,
\end{eqnarray}
where \( \omega = 2 v \cdot k \) is twice the residual off-shell energy. Here, we project out the Lorentz structure associated with the vector meson contribution, while subleading terms are omitted for brevity.

At the hadronic level, Equation~(\ref{eq:pi}) can be rewritten as
\begin{equation}
\Pi(\omega) = \frac{f^{2}}{2\overline{\Lambda} - \omega} + \text{higher states} \, ,
\label{eq:pole}
\end{equation}
where
\begin{equation}
\overline{\Lambda} \equiv \lim_{m_t \rightarrow \infty} (m_{T_s^\star} - m_t) \, ,
\label{eq:leading}
\end{equation}
denotes the residual mass of the \( T_s^\star \) meson in the heavy quark limit, with \( m_{T_s^\star} \) representing its physical mass.

On the QCD side, Equation~(\ref{eq:pi}) is evaluated via the operator product expansion (OPE). By substituting the interpolating current defined in Equation~(\ref{def:J1}) and performing a Borel transformation, we arrive at
\begin{align}
\Pi(\omega_c, T) &= f^2(\omega_c, T) \cdot e^{-2\overline{\Lambda}(\omega_c, T) / T} \nonumber\\
&= \int_{s_<}^{\omega_c} \left[\frac{3\omega^{2}}{16\pi^2}
+\frac{3m_s\omega}{8\pi^2}-\frac{3m_s^2}{8\pi^2} \right]
e^{-{\omega / T}} \, d\omega 
\nonumber\\
&- \frac{\langle\bar{s}s\rangle}{2}+\frac{m_s \langle\bar{s}s\rangle}{4T}+\frac{\langle g_s \bar{s} \sigma G s \rangle}{8T^2} \, .
\label{eq:ope}
\end{align}
{Here,} 
 \( s_< = 2m_s \) represents the physical threshold, \( \omega_c \) is the continuum threshold parameter, and \( T \) denotes the Borel mass. The parameter \( \omega_c \) encodes the quark–hadron duality assumption, wherein contributions from higher excited states and the continuum are modeled by the perturbative portion of the OPE. In practice, \( \omega_c/2 \) is selected to lie moderately above the residual mass of the meson under study, thereby ensuring ground-state dominance and reliable convergence of the OPE expansion.

{When calculating Equation~(\ref{eq:ope}), we include the perturbative contribution, the strange quark mass \( m_s \), the quark condensate \( \langle \bar{s} s \rangle \), and the quark–gluon mixed condensate \( \langle g_s \bar{s} \sigma G s \rangle \). Here, \( g_s \) denotes the strong coupling constant, and \( G_{\mu\nu}^a \) is the gluon field strength tensor. The mixed condensate is explicitly defined as
\[
\langle g_s \bar{s} \sigma G s \rangle \equiv \left\langle g_s\, \bar{s}_i\, \sigma^{\mu\nu} \left( \frac{\lambda^a}{2} \right)_{ij} G_{\mu\nu}^a\, s_j \right\rangle,
\]
where \( i, j \) are color indices, and \( \lambda^a \) are the Gell--Mann matrices of $SU(3)$.}

To extract physical observables, we differentiate Equation~(\ref{eq:ope}) with respect to \( -2/T \), yielding
\begin{eqnarray}
\overline{\Lambda}(\omega_c, T)
&=& \frac{1}{\Pi(\omega_c, T)} \cdot \frac{\partial \Pi(\omega_c, T)}{\partial(-2/T)} \, ,
\label{eq:mass}
\\
f^2(\omega_c, T)
&=& \Pi(\omega_c, T) \cdot e^{2\overline{\Lambda}(\omega_c, T) / T} \, .
\label{eq:coupling}
\end{eqnarray}
{These} expressions involve two free parameters: the threshold value \( \omega_c \) and the Borel mass \( T \). To determine their optimal working ranges, we impose three standard QCD sum rule criteria. Throughout our numerical analysis, we adopt the following input parameters at the renormalization scale \( \mu = 2~\mathrm{GeV} \)~\citep{pdg,Yang:1993bp,Narison:2011xe,Narison:2018dcr,Gimenez:2005nt,Jamin:2002ev,Ioffe:2002be,Ovchinnikov:1988gk,Colangelo:2000dp}:
\begin{eqnarray}
\nonumber \langle \bar{q} q \rangle &=& -(0.240~\mathrm{GeV})^3 \, , 
\\ \nonumber \langle \bar{s} s \rangle &=& (0.8 \pm 0.1) \times \langle \bar{q} q \rangle \, , 
\\ \nonumber \langle g_s \bar{q} \sigma G q \rangle &=& M_0^2 \cdot \langle \bar{q} q \rangle \, , 
\\ \langle g_s \bar{s} \sigma G s \rangle &=& M_0^2 \cdot \langle \bar{s} s \rangle \, ,
\label{eq:condensate}
\\ \nonumber M_0^2 &=& (0.8 \pm 0.2)~\mathrm{GeV}^2 \, , 
\\ \nonumber \langle g_s^2 GG \rangle &=& (0.48\pm 0.14) ~\mathrm{GeV}^4 \, , 
\\ \nonumber m_s &=& 93.5 \pm 0.8~\mathrm{MeV} \, ,
\\ \nonumber m_q &\approx& m_u \approx m_d \approx 0~\mathrm{MeV} \, .
\end{eqnarray}

\begin{figure*}[htbp]
\centering
\subfigure[]{\scalebox{0.42}{\includegraphics{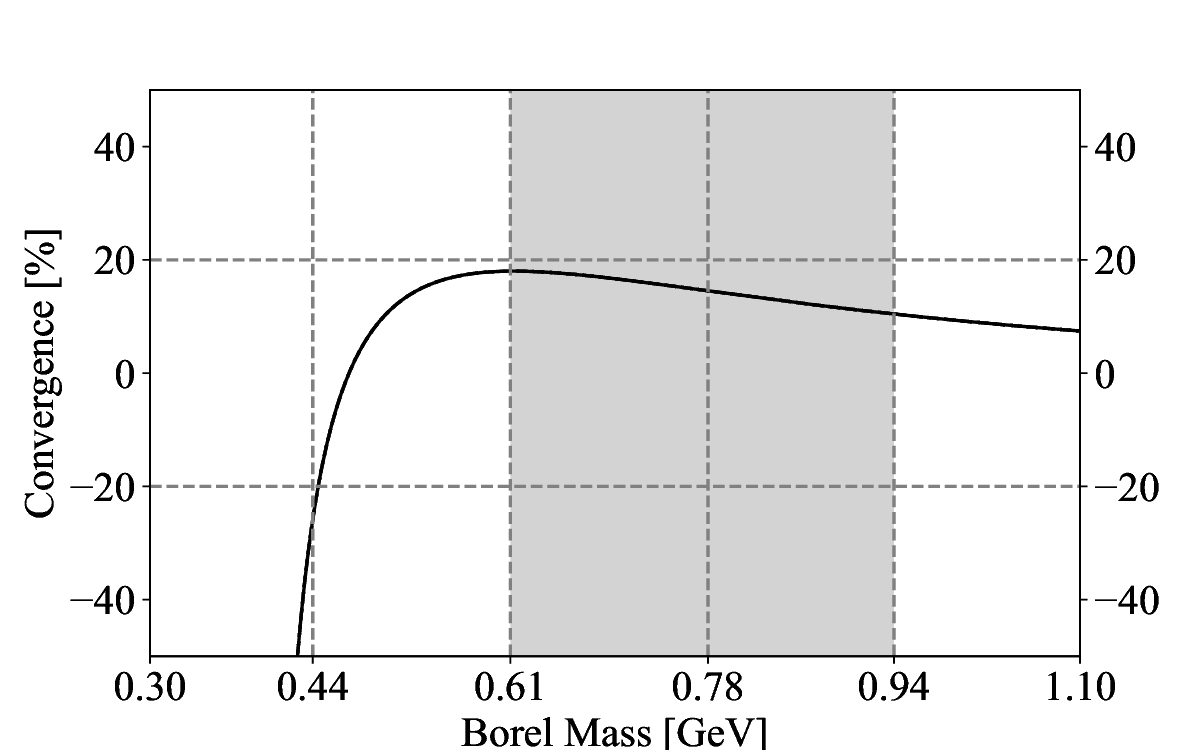}}}
\hspace{0.5cm}
\subfigure[]{\scalebox{0.42}{\includegraphics{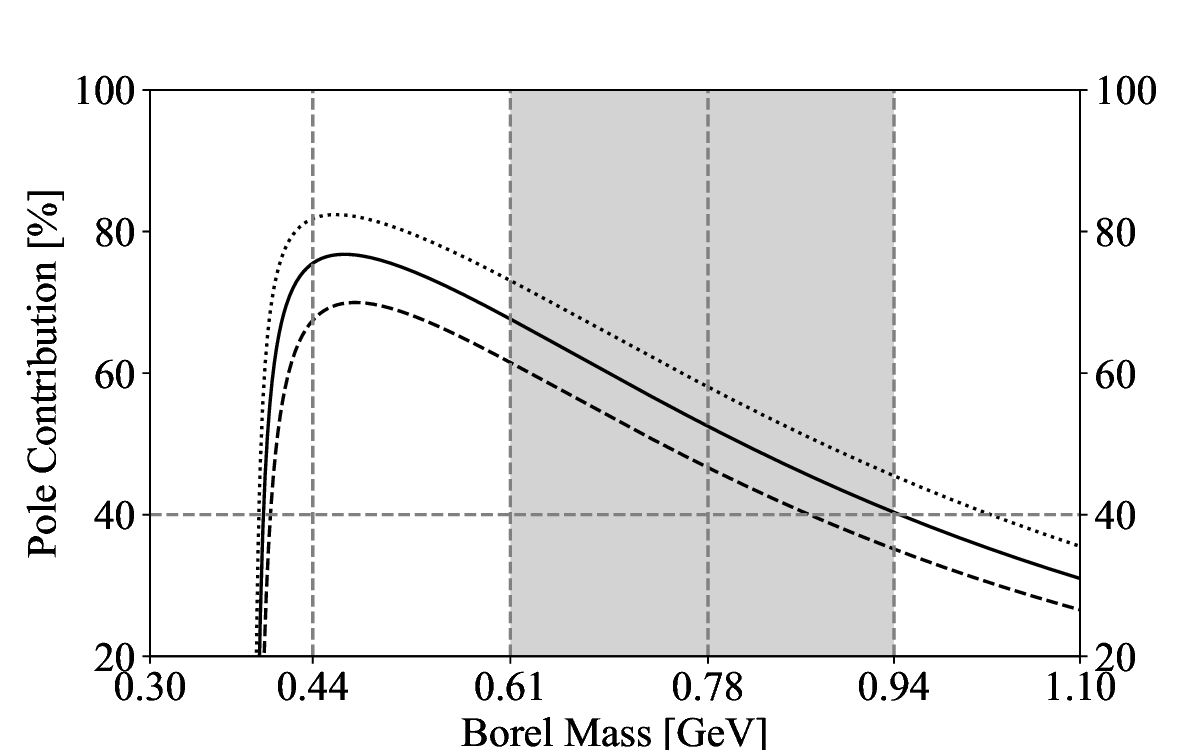}}}
\caption{Dependence of (a) the convergence parameter (CVG), defined in Eq.~(\ref{eq_convergence}), and (b) the pole contribution (PC), defined in Eq.~(\ref{eq:pole}),  for the interpolating current \( J_\mu(x) \). The Borel window \( 0.61~\mathrm{GeV} < T < 0.94~\mathrm{GeV} \) is indicated by the gray-shaded region. Dashed, solid, and dotted curves correspond to \( \omega_c = 1.65 \), \( 1.85 \), and \( 2.05~\mathrm{GeV} \), respectively.}
\label{fig:pole}
\end{figure*}

The first criterion requires that contributions from higher-dimensional condensates be less than 20\% of the total correlator:
\begin{equation}
\label{eq_convergence}
\mathrm{CVG} \equiv \left| \frac{ \Pi^{\text{high-order}}(\omega_c, T) }{ \Pi(\infty, T) } \right| \leq 20\% \, .
\end{equation}
{As} shown in Figure~\ref{fig:pole}a, this condition sets the lower bound of the Borel mass at \mbox{\( T_{\rm min} = 0.44~\mathrm{GeV} \)}. 

\begin{figure*}[htbp]
\centering
\subfigure[]{\scalebox{0.42}{\includegraphics{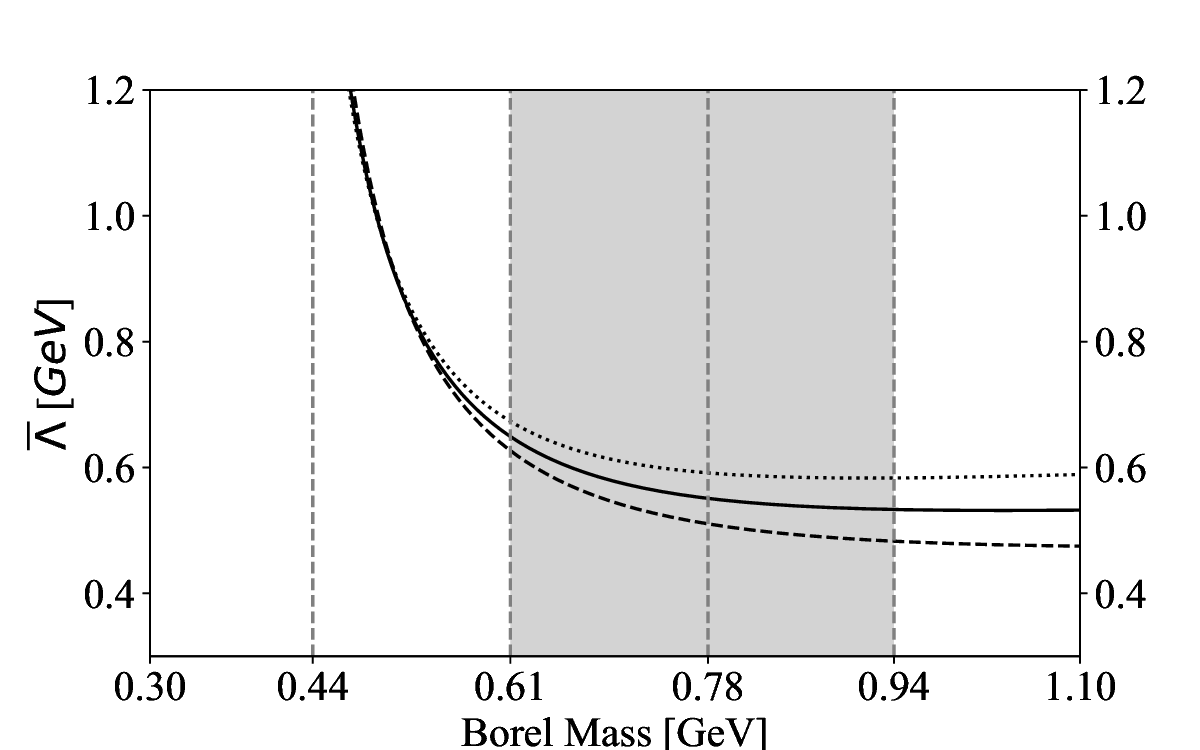}}}
\hspace{0.5cm}
\subfigure[]{\scalebox{0.42}{\includegraphics{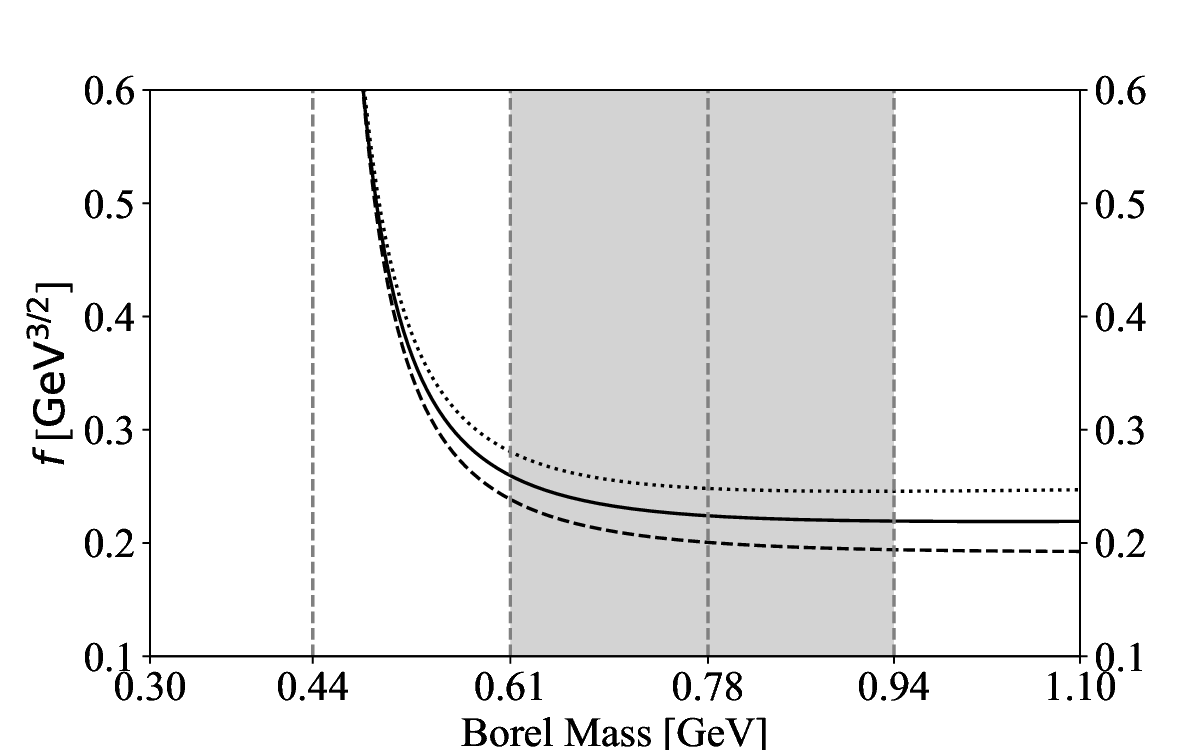}}}
\caption{Variations of (a) the residual mass \( \overline{\Lambda} \) and (b) the decay constant \( f \) with respect to the Borel mass \( T \) for the interpolating current \( J_\mu(x) \). The Borel window \( 0.61~\mathrm{GeV} < T < 0.94~\mathrm{GeV} \) is indicated by the gray-shaded region. Dashed, solid, and dotted curves correspond to \( \omega_c = 1.65 \), \( 1.85 \), and \( 2.05~\mathrm{GeV} \), respectively.}
\label{fig:leading}
\end{figure*}

The second criterion demands that the pole contribution exceeds 40\%:
\begin{equation}
\label{eq_pole}
\mathrm{PC} \equiv \frac{ \Pi(\omega_c, T) }{ \Pi(\infty, T) } \geq40\% \, .
\end{equation}
{As} illustrated in Figure~\ref{fig:pole}b, this requirement establishes an upper limit \( T_{\rm max} = 0.94~\mathrm{GeV} \) for \( \omega_c = 1.85~\mathrm{GeV} \). Together, these two conditions define the initial Borel window:
\begin{equation}
0.44~\mathrm{GeV} < T < 0.94~\mathrm{GeV} \, .
\end{equation}
{To} further refine this window, we analyze the Borel-mass dependence of the extracted parameters \( \overline{\Lambda} \) and \( f \), as shown in Figure~\ref{fig:leading}. The convergence ratio CVG reaches a local maximum near \( T_{\rm peak} = 0.61~\mathrm{GeV} \); beyond this value, both extracted observables exhibit significantly improved stability. In contrast, for \( T < T_{\rm peak} \), the strong variation in \( \overline{\Lambda} \) and \( f \) with respect to \( T \) violates the Borel stability condition. Consequently, the third criterion restricts the Borel window to
\begin{equation}
0.61~\mathrm{GeV} < T < 0.94~\mathrm{GeV} \, .
\end{equation}
{Within} this optimized range, we extract the following numerical results:
\begin{eqnarray}
\overline{\Lambda} &=& 0.55^{+0.12}_{-0.07}~\mathrm{GeV} \, , \\
f &=& 0.22^{+0.04}_{-0.03}~\mathrm{GeV}^{3/2} \, ,
\end{eqnarray}
where the central values correspond to \( T = 0.78~\mathrm{GeV} \) and \( \omega_c = 1.85~\mathrm{GeV} \).

\section{Results at the ${\mathcal O}(1/m_Q)$ Order}
\label{sec:nexttoleading}

In this section we extend our analysis to incorporate \( \mathcal{O}(1/m_Q) \) corrections~\citep{Dai:1996qx,Dai:2003yg}. To this end, we adopt the effective Lagrangian of HQET:
\begin{eqnarray}
\mathcal{L}_{\rm eff} = \bar{h}_{v}^a\, i v \cdot D\, h_{v}^a + \frac{1}{2m_t} \mathcal{K} + \frac{1}{2m_t} \mathcal{S} \, ,
\label{eq:next}
\end{eqnarray}
where \( \mathcal{K} \) is the nonrelativistic kinetic energy operator,
\begin{eqnarray}
\mathcal{K} = \bar{h}_{v}^a (i D_{t})^{2} h_{v}^a \, ,
\end{eqnarray}
and \( \mathcal{S} \) is the Pauli term responsible for the chromomagnetic interaction:
\begin{eqnarray}
\mathcal{S} = \frac{g_s}{2} C_{\rm mag} \left( \frac{m_t}{\mu} \right) \bar{h}_{v}^a \sigma_{\mu\nu} G^{\mu\nu} h_{v}^a \, .
\end{eqnarray}
{The} corresponding Wilson coefficient is given by
\begin{eqnarray}
C_{\rm mag} \left( \frac{m_t}{\mu} \right) = \left[ \frac{\overline{\alpha}_s(m_t)}{\overline{\alpha}_s(\mu)} \right]^{3/\beta_0} \, ,
\end{eqnarray}
where \( \beta_0 = 11 - \frac{2}{3} n_f \). At the renormalization scale \( \mu = 2~\mathrm{GeV} \), this evaluates to \mbox{\( C_{\rm mag}(\mu) \approx 0.6 \)} for the top quark.

At the hadronic level, the pole term in the correlation function can be expanded to include \( \mathcal{O}(1/m_Q) \) corrections:
\begin{align}
\Pi(\omega) &= \frac{(f + \delta f)^2}{2(\overline{\Lambda} + \delta m) - \omega}
\label{eq:correction}
\nonumber\\
&= \frac{f^2}{2\overline{\Lambda} - \omega} 
- \frac{2\delta m\, f^2}{(2\overline{\Lambda} - \omega)^2} 
+ \frac{2f\, \delta f}{2\overline{\Lambda} - \omega} \, ,
\end{align}
where \( \delta m \) and \( \delta f \) denote the \( \mathcal{O}(1/m_Q) \) corrections to the hadron mass \( m_{T_s^\star} \) and the decay constant \( f \), respectively.

To evaluate the mass correction \( \delta m \), we consider the three-point correlation functions:
\begin{eqnarray}
&& \delta_O \Pi_{\mu\nu}(\omega, \omega^\prime)
\label{eq:nextpi}
\\ \nonumber &=& i^2 \int d^4x\, d^4y\, e^{i k \cdot x - i k^\prime \cdot y} \langle 0 | T[J_\mu(x)\, O(0)\, J_\nu^\dagger(y)] | 0 \rangle
\\ \nonumber &=& \left(g_{\mu\nu} - \frac{q_\mu q_\nu}{q^2} \right) \times \delta_O \Pi(\omega, \omega^\prime) + \cdots \, ,
\end{eqnarray}
where \( O = \mathcal{K} \) or \( \mathcal{S} \). According to the effective Lagrangian in Equation~(\ref{eq:next}), the hadronic representations of these correlators are given by
\begin{eqnarray}
\delta_{\mathcal{K}} \Pi(\omega, \omega^\prime) &=& \frac{f^2 K}{(2\overline{\Lambda} - \omega)(2\overline{\Lambda} - \omega^\prime)} + \cdots \, ,
\label{eq:K}
\\
\delta_{\mathcal{S}} \Pi(\omega, \omega^\prime) &=& \frac{d_M f^2 \Sigma}{(2\overline{\Lambda} - \omega)(2\overline{\Lambda} - \omega^\prime)} + \cdots \, ,
\label{eq:S}
\end{eqnarray}
with the matrix elements defined as
\begin{eqnarray}
K &\equiv& \langle T_s^\star | \bar{h}_v^a (i D_\perp)^2 h_v^a | T_s^\star \rangle \, ,
\\
d_M \Sigma &\equiv& \left\langle T_s^\star \left| \frac{g_s}{2} \bar{h}_v^a \sigma_{\mu\nu} G^{\mu\nu} h_v^a \right| T_s^\star \right\rangle \, .
\end{eqnarray}
{The} spin-dependent coefficient \( d_M \) is given by
\begin{eqnarray}
d_M &=& d_{j,j_l} \, ,
\\
d_{j_l - 1/2, j_l} &=& 2 j_l + 2 \, ,
\\
d_{j_l + 1/2, j_l} &=& -2 j_l \, .
\end{eqnarray}
{By} setting \( \omega = \omega^\prime \) and comparing Equations~(\ref{eq:correction}), (\ref{eq:K}), and (\ref{eq:S}), the mass correction is \mbox{obtained as}
\begin{eqnarray}
\delta m = -\frac{1}{4 m_t} \left( K + d_M C_{\rm mag} \Sigma \right) \, .
\label{eq:more}
\end{eqnarray}

\begin{figure*}[hbt]
\centering
\subfigure[]{\scalebox{0.42}{\includegraphics{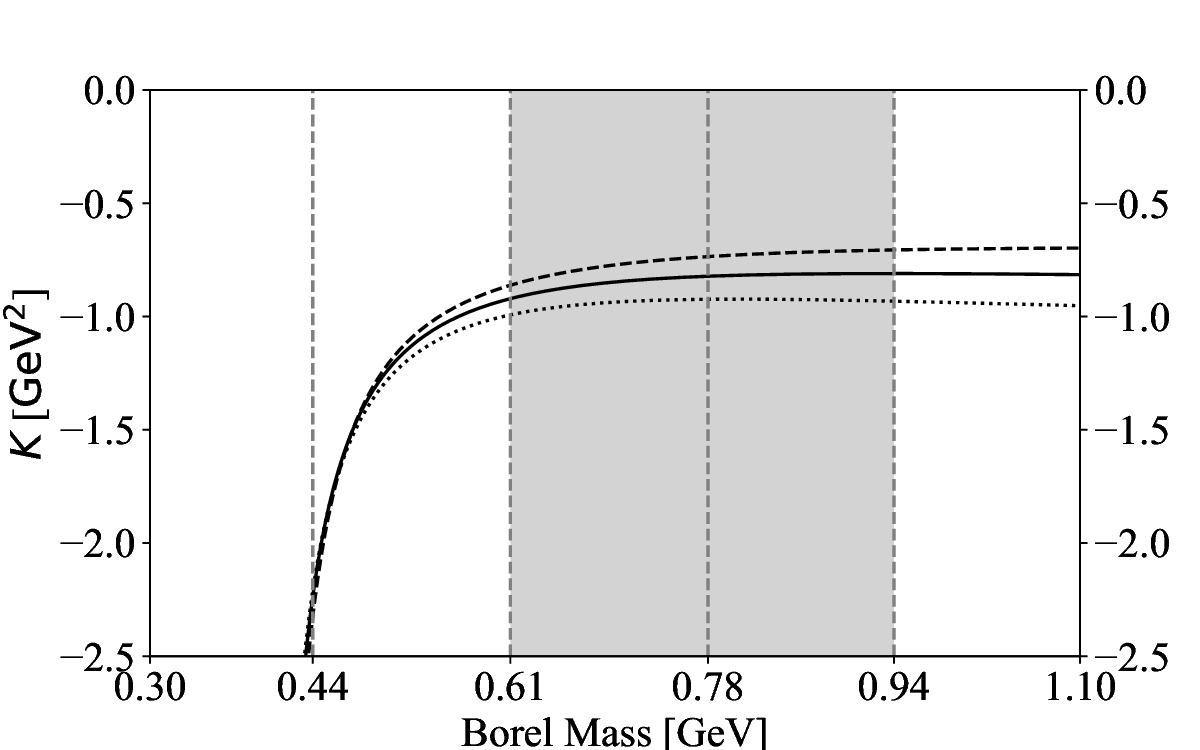}}}
~~~~~
\subfigure[]{\scalebox{0.42}{\includegraphics{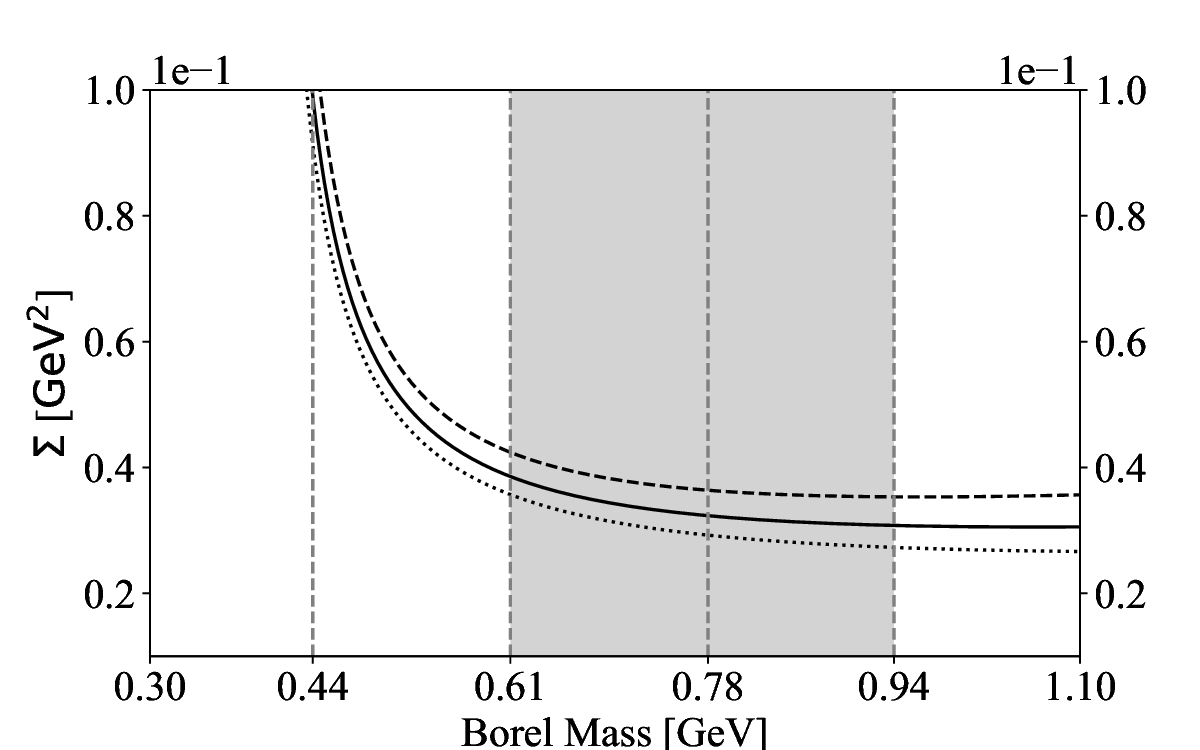}}}
\caption{Dependence of (a) the kinetic matrix element \( K \) and (b) the chromomagnetic matrix element \(  \Sigma \) on the Borel mass \( T \), calculated using the interpolating current \( J_\mu(x) \). The Borel window \( 0.61~\mathrm{GeV} < T < 0.94~\mathrm{GeV} \) is indicated by the gray-shaded region. Dashed, solid, and dotted curves correspond to continuum thresholds \( \omega_c = 1.65 \), \( 1.85 \), and \( 2.05~\mathrm{GeV} \), respectively.}
\label{fig:KandS}
\end{figure*}

The correlation functions in Equation~(\ref{eq:nextpi}) can also be evaluated on the QCD side using the operator product expansion (OPE). By inserting Equation~(\ref{def:J1}) into Equation~(\ref{eq:nextpi}) and applying a double Borel transformation with respect to \( \omega \) and \( \omega^\prime \), we obtain the corresponding sum rules involving two Borel parameters, \( T_1 \) and \( T_2 \). For simplicity, we set \( T_1 = T_2 = T \), leading to
\begin{eqnarray}
&& f^2 K(\omega_c, T) \, e^{-2\overline{\Lambda} / T}
\label{eq:Kc}
\\ \nonumber &=&\int_{s_<}^{\omega_c}
\left[-\frac{3 \omega^4}{32 \pi^2}
-\frac{3 m_s \omega^3}{16 \pi^2}
+\frac{9 m_s^2 \omega^2}{16 \pi^2}
\right] e^{-{\omega / T}} d\omega
\\ \nonumber && + \frac{3\langle g_s \bar{s} \sigma G s \rangle}{8} \, ,
\\
&& f^2 \Sigma(\omega_c, T)\, e^{-2\overline{\Lambda} / T}
= \frac{\langle g^2 GG \rangle }{96\pi^2} T\, .
\label{eq:Sc}
\end{eqnarray}
{The} Borel mass dependence of \( K \) and \( \Sigma \) is shown in Figure~\ref{fig:KandS}. Both quantities exhibit mild variation within the Borel window \( 0.61~\mathrm{GeV} < T < 0.94~\mathrm{GeV} \). Within this range, we extract the following numerical results:
\begin{eqnarray}
K &=& -0.82^{+0.13}_{-0.17}~\mathrm{GeV}^2 \, ,
\\
\Sigma &=& 0.032^{+0.012}_{-0.010}~\mathrm{GeV}^2 \, .
\end{eqnarray}
\vspace{-24pt} 

\section{Discussion and Conclusions}
\label{sec:summary}

Based on the results obtained in Sections~\ref{sec:leading} and~\ref{sec:nexttoleading}, we now determine the mass of the vector topped meson \( T_s^\star \). A key aspect of this analysis is the choice of renormalization scheme for the top quark mass. In our previous QCD sum rule studies of charmed and bottom baryons~\cite{Chen:2015kpa,Mao:2015gya}, the $\overline{\mathrm{MS}}$ scheme has been commonly employed:
\begin{eqnarray}
m_c^{\overline{\mathrm{MS}}} &=& 1.2730^{+0.0046}_{-0.0046}~\mathrm{GeV} \, , 
\\
m_b^{\overline{\mathrm{MS}}} &=& 4.183^{+0.007}_{-0.007}~\mathrm{GeV} \, ,
\end{eqnarray}
with corresponding pole masses,
\begin{eqnarray}
m_c^{\mathrm{pole}} &=& 1.67^{+0.07}_{-0.07}~\mathrm{GeV} \, , 
\\
m_b^{\mathrm{pole}} &=& 4.78^{+0.06}_{-0.06}~\mathrm{GeV} \, .
\end{eqnarray}
{These} differences are moderate, typically within several hundred MeV.

However, for the top quark, the discrepancy between the two schemes becomes significantly larger:
\begin{eqnarray}
m_t^{\overline{\mathrm{MS}}} &=& 162.5^{+2.1}_{-1.5}~\mathrm{GeV} \, , 
\\
m_t^{\mathrm{pole}} &=& 172.57^{+0.29}_{-0.29}~\mathrm{GeV} \, ,
\end{eqnarray}
indicating a gap of nearly \( 10~\mathrm{GeV} \). Since the residual mass \( \overline{\Lambda} \) is only of order \( 1~\mathrm{GeV} \), such a large deviation would compromise the numerical reliability of the extracted meson mass. Therefore, we adopt the pole mass of the top quark in our final evaluation of the \( T_s^\star \) meson mass. This choice also suggests that pole masses might generally be more suitable in QCD sum rule applications—a possibility that merits further investigation. Additionally, when electroweak corrections are considered, the 
$\overline{\text{MS}}$ mass of the top quark approaches its pole mass. For further details, the reader is referred to Ref.~\citep{Kataev:2022dua}.

Using Equations~(\ref{eq:leading}) and~(\ref{eq:more}), the mass of the ground-state \( T_s^\star \) meson is given by
\begin{eqnarray}
m_{T_s^\star} 
&=& m_t + \overline{\Lambda} + \delta m 
\\ \nonumber 
&=& \left[ 172.57^{+0.29}_{-0.29} 
+ 0.55^{+0.12}_{-0.07} \right]~\mathrm{GeV} 
+ 1.2^{+0.3}_{-0.2}~\mathrm{MeV}
\\ \nonumber 
&=& 173.12^{+0.31}_{-0.30}~\mathrm{GeV} \, .
\end{eqnarray}
{{The} quoted uncertainty stems from the pole mass of the top quark, the two QCD sum rule parameters—the continuum threshold \( \omega_c \) and the Borel mass \( T \)—and the QCD inputs listed in Equation~(\ref{eq:condensate}). We note that the uncertainty associated with the choice of mass scheme—whether to use the top quark pole mass or its $\overline{\mathrm{MS}}$ value—is not included here. Its propagation would induce a non-negligible additional error in the predicted topped-meson masses; {\it e.g.}, using the $\overline{\mathrm{MS}}$ scheme yields \( m_{T_s} = 163.0^{+2.1}_{-1.5} \)~GeV and \( m_{T_s^\star} = 163.1^{+2.1}_{-1.5} \)~GeV. This constitutes an additional systematic uncertainty to be addressed in future work.}

In addition to the \( T_s^\star \) meson, other ground-state topped mesons have been analyzed using similar methods, with the results summarized in Table~\ref{tab:result}. In general, the masses of these mesons lie around \( 173.1~\mathrm{GeV} \), approximately $0.5$–\( 0.6~\mathrm{GeV} \) above the top quark’s pole mass. For comparison, ground-state bottom mesons are typically located about \( 0.5 \)–\( 0.6~\mathrm{GeV} \) above the bottom quark pole mass, whereas charmed mesons usually lie \( 0.2 \)–\( 0.4~\mathrm{GeV} \) above the charm quark pole mass~\citep{pdg}.

\begin{table*}[hbtp]
\begin{center}
\renewcommand{\arraystretch}{1.8}
\caption{Parameters of ground-state topped mesons calculated using QCD sum rules within the HQET framework. The states \( T^{(\star)} \) and \( T_s^{(\star)} \) contain the quark contents \( t \bar{q} \) (\( \bar q = \bar u/ \bar d \)) and \( t \bar{s} \), respectively. The last column lists several possible decay channels.}
\begin{tabular}{ c | c | c | c | c | c | c | c | c}
\hline\hline
\multirow{2}{*}{~~~~} & $\omega_c$& ~~~~~Working region~~~~~ & ~~~~~~$\overline{\Lambda}$~~~~~~ & ~~Meson~~ & ~~~~~Mass~~~~~~ & ~Difference~ & ~~~~~~~$f$~~~~~~~ & \multirow{2}{*}{~Decay channels~}
\\                                               & (GeV) & (GeV)      & (GeV)                &                               ($J^P$)       & (GeV)      & (MeV)        & (GeV$^{3/2}$) & 
\\ \hline\hline
\multirow{2}{*}{$T^{(*)}$}& \multirow{2}{*}{1.65}& \multirow{2}{*}{$0.79\le T \le0.89$}& \multirow{2}{*}{$0.46 _{-0.07}^{+0.07}$}& $T~(0^-)$& $173.03_{-0.30}^{+0.30}$& \multirow{2}{*}{$0.12_{-0.04}^{+0.04}$}&\multirow{2}{*}{ $0.19_{-0.02}^{+0.02}$ }& 
$\Upsilon D^{(*)}$, $\bar B_c^{(*)} B^{(*)}$,
\\ \cline{5-6}& & & & $T^\star~(1^-)$& $173.03_{-0.30}^{+0.30}$&& & 
$B_s^{(*)} D^{(*)}$, $D_s^{(*)} B^{(*)}$
\\ \hline
\multirow{2}{*}{$T_s^{(*)}$}& \multirow{2}{*}{1.85}& \multirow{2}{*}{$0.61\le T \le 0.94$}& \multirow{2}{*}{$0.55 _{-0.07}^{+0.12}$}& $T_s~(0^-)$& $173.12_{-0.30}^{+0.31}$& \multirow{2}{*}{$0.11_{-0.04}^{+0.04}$}& \multirow{2}{*}{$0.22_{-0.03}^{+0.04}$ }& 
$\Upsilon D_s^{(*)}$, $\bar B_c^{(*)} B_s^{(*)}$, \\ \cline{5-6}& & & & $T_s^\star~(1^-)$& $173.12_{-0.30}^{+0.31}$&&&
$B_s^{(*)} D_s^{(*)}$
\\ \hline\hline
\end{tabular}
\label{tab:result}
\end{center}
\end{table*}

In summary, this work presents the first application of QCD sum rules within the HQET framework to mesonic systems containing a top quark. In parallel, singly topped baryons are studied in Ref.~\citep{Zhang:2025xxd} using the same formalism. Although both topped mesons and singly topped baryons are challenging to form due to the extremely short lifetime of the top quark, their theoretical study offers valuable insights into QCD dynamics in the heavy quark limit. These systems are expected to inherit the top quark decay width, \( \Gamma_t \approx 1.41\,\mathrm{GeV} \)~\citep{pdg}, which is comparable to that of the shortest-lived hadronic resonances, such as the \( \sigma(500) \) and \( \kappa(800) \). The decay widths of these resonances are estimated to be in the range of \( 400 \)–\( 700~\mathrm{MeV} \) for the \( \sigma(500) \) and \( 500 \)–\( 800~\mathrm{MeV} \) for the \( \kappa(800) \)~\citep{pdg}. This places topped hadrons in a transitional regime where weak decay competes directly with strong binding, providing a unique opportunity to probe the onset of hadron formation at the extreme heavy quark mass scale~\cite{Chen:2023zlj}. 
{ While the results of this study are speculative and exploratory, they provide an important step toward understanding the behavior of topped hadrons. Experimental verification of these predictions could either challenge the methods used in this study or offer insight into why they work in this context. In either case, the results will provide valuable information for further theoretical investigations into the nature of these novel hadronic states.}

\section*{Acknowledgments}

This work was supported by the National Natural Science Foundation of China under Grant No.~12075019,  
the Jiangsu Provincial Double-Innovation Program under Grant No.~JSSCRC2021488,  
the China Postdoctoral Science Foundation under Grant No.~2024M750049,  
the SEU Innovation Capability Enhancement Plan for Doctoral Students,  
and 
the Fundamental Research Funds for the Central Universities.

\bibliographystyle{elsarticle-num}
\bibliography{ref}

\end{document}